\documentclass[12pt]{iopart}
\usepackage{graphicx}
\usepackage{dsfont}
\usepackage[T1]{url}

\providecommand{\sunmass}{\mathrm{M}_{\scriptscriptstyle\odot}}
\providecommand{\prob}{\mathrm{P}}
\providecommand{\realline}{\mathds{R}}

\begin{document}
\title{Coherent Bayesian analysis of inspiral signals}
 \author{Christian~R\"over$^1$,
         Renate~Meyer$^1$,
         Gianluca~M.~Guidi$^2$,
         Andrea~Vicer\'e$^2$ and 
         Nelson~Christensen$^3$}
 \address{$^1$ Department of Statistics, The University of Auckland,
          Auckland, New Zealand}
 \address{\mbox{$^2$ Universit\`a degli Studi di Urbino ``Carlo~Bo'', Urbino, Italy,}
          \mbox{\phantom{$^2$} and INFN, Sezione di Firenze, Florence, Italy}}
 \address{$^3$ Physics and Astronomy, 
          Carleton College, Northfield, MN, USA}

\begin{abstract} 
We present in this paper a Bayesian parameter estimation method 
for the analysis of interferometric gravitational wave observations 
of an inspiral of binary compact objects using data recorded 
simultaneously by a network of several interferometers at different sites. 
We consider neutron star or black hole inspirals that are modeled 
to 3.5~post-Newtonian (PN) order in phase and 2.5~PN in amplitude.
Inference is facilitated using Markov chain Monte Carlo methods 
that are adapted in order to efficiently explore 
the particular parameter space. 
Examples are shown to illustrate how and what information 
about the different parameters can be derived from the data. 
This study uses simulated signals and data with noise characteristics 
that are assumed to be defined by the LIGO and Virgo detectors 
operating at their design sensitivities.
Nine parameters are estimated, including those associated 
with the binary system, plus its location on the sky. 
We explain how this technique will be part of a detection pipeline 
for binary systems of compact objects with masses up to~$20\;\sunmass$,
including cases where the ratio of the individual masses can be extreme.
\end{abstract}


\submitto{Classical and Quantum Gravity}

\section{Introduction}
A world-wide network of interferometric gravitational wave detectors 
is now on-line. 
LIGO has reached its target sensitivity~\cite{AbbottEtAl2004a,Sigg2004}, 
and Virgo is fast approaching theirs~\cite{AcerneseEtAl2006,AcerneseEtAl2005}. 
GEO~\cite{LueckEtAl2006} and TAMA~\cite{TakahashiEtAl2003} are also 
participating in the search for gravitational waves. 
Compact binary systems will certainly produce 
gravitational waves~\cite{TaylorWeisberg1989}, 
and they are likely to be one of the most promising sources.

The LIGO Scientific Collaboration (LSC)~\cite{AbbottEtAl2004b} 
and Virgo~\cite{MarionEtAl2003,AmicoEtAl2003} each have 
search pipelines for binary inspiral events, 
and studies have shown that these pipelines have equivalent 
detection capabilities~\cite{BeauvilleEtAl2007}.
The LSC has conducted searches for binary neutron star 
inspirals~\cite{AbbottEtAl2004b,AbbottEtAl2005a},
primordial black hole binary coalescences 
in the galactic halo~\cite{AbbottEtAl2005b}, 
and black hole binaries~\cite{AbbottEtAl2006a}. 
The LSC and TAMA have conducted a joint search 
for binary neutron star systems~\cite{AbbottEtAl2006b}, 
and soon the LSC and Virgo will be conducting 
collaborative searches~\cite{BeauvilleEtAl2007}.

The purpose of a binary inspiral detection pipeline is to find 
a signal within the data. 
Once researchers suspect that a signal is present then parameter 
estimation techniques can be applied in order to produce estimates 
and summary statistics for the astrophysical parameters. 
Bayesian Markov chain Monte Carlo (MCMC) methods~\cite{MCMCinPractice}
are well suited for this problem, especially since it is possible 
to produce accurate predictions for the form of the signal. 
MCMC parameter estimation techniques have been developed 
for binary neutron star inspirals, 
as seen by a single interferometer~\cite{RoeverMeyerChristensen2006a}. 
In addition, MCMC methods have been developed for the coherent analysis 
of data from a world-wide network of 
interferometers~\cite{RoeverMeyerChristensen2007a}.

A difficult detection scenario involves finding a signal produced 
by a binary system where the mass ratio between the two objects is large. 
In such a case, the signal will likely have its amplitude significantly 
modulated (as opposed to just a `simple' chirp with monotonously increasing
frequency and amplitude), 
and it will be necessary to use 
higher-order post-Newtonian (PN) approximations.
In this paper we present a description of our method for producing 
parameter estimates associated with a binary inspiral modeled 
to 3.5~post-Newtonain (PN) order in phase, 
and 2.5~PN 
\nocite{BlanchetEtAl2002,BlanchetEtAl2005a,ArunEtAl2004,ArunEtAl2005}
in amplitude~\cite{BlanchetEtAl2002,ArunEtAl2004}.
There are numerous goals that we wish to address 
with this version of our code. 
We employ new and more advanced MCMC methods, such as 
evolutionary MCMC~\cite{LiangWong2001a}.
The higher order PN templates will also allow for examination 
of signals where the amplitude is modulated, 
as may be the case with rather large ratios between 
the masses of the compact objects.
Finally, we see this MCMC program as part of a larger detection pipeline 
for signals from binary inspirals with large mass ratios, and 
individual masses going up to $20\;\sunmass$. 
We imagine, for example, using an existing de\-tec\-tion 
pipeline~\cite{AmicoEtAl2003} to generate a reasonable number of triggers; 
the MCMC would then analyze each of the triggers in detail. 
Once the MCMC has reached convergence, 
an estimate for the signal parameters would be produced.  
In this paper we provide a description of the MCMC component 
of this detection pipeline. 

\section{Inference framework}
\subsection{The Bayesian approach}
We follow a Bayesian approach in order to do inference 
on the inspiral signal's parameters, 
since this better allows one to address the questions 
of immediate interest in such a context.
Other methods (e.g. matched-filtering methods)
on the other hand usually follow a Maximum-Likelihood approach, 
which does not yield as satisfactorily interpretable results, 
and does not exploit the information available in the data 
to the same extent
\cite{Finn1997}.
In a Bayesian setup, information about parameters is fomulated in terms
of probability distributions on the parameter space. 
First, the pre-observational knowledge is expressed 
in the \textsl{prior distribution}, and inference eventually is done 
through the parameters' \textsl{posterior distribution} 
that is conditional on the observed data,
and follows through the application of Bayes' theorem
on the prior and the data model (likelihood).
The parameters' posterior distribution then expresses 
the information about the parameters given the prior knowledge,
the model, and the data at hand~\cite{Jaynes,Gregory,BDA}.

\subsection{MCMC methods}
Once the Bayesian framework is set up, 
inference depends on evaluating the parameters' posterior distribution,
which is given in terms of the (non-normalized) posterior density, 
in our case
a function of 9~parameters.
Typically, one will be interested in figures such as 
posterior means, confidence bounds,
or marginal densities for individual parameters, 
which requires integration of the posterior over the parameter space.
This problem is commonly approached using Monte Carlo integration, 
i.e.\ by simulating random draws from the posterior distribution, 
and then approximating the desired integrals by sample statistics
(means by averages, etc.).
The most popular algorithms for this purpose are 
Markov chain Monte Carlo (MCMC) samplers
that simulate a random walk through parameter space 
whose stationary distribution is the 
posterior distribution~\cite{BDA,MCMCinPractice}.

Metropolis- (and related) MCMC algorithms also have nice optimization
properties. In fact they happen to behave very similar to e.g.\
a Nelder-Mead algorithm which is extended to a simulated annealing
algorithm; on its random walk through parameter space it will
always accept an `uphill' step, and sometimes (randomly) 
a `downhill' step as well~\cite{NumRecipes}.
This property often comes in handy since the 
problem---as in our case---usually 
is not only to \textsl{sample} from the posterior, 
but also to first \textsl{find} the global posterior mode(s) within a
complex posterior surface, and among numerous minor modes.
These convergence properties can also be enhanced 
through the implementation of the sampler,
while care must be taken to maintain its ergodicity properties.

For our purposes we used a basic Metropolis sampler that 
we recently upgraded to an \textsl{evolutionary MCMC} algorithm~\cite{LiangWong2001a}, 
a generalization that is motivated by genetic algorithms~\cite{Goldberg}.
This extension offers substantial improvement over the previously
employed parallel tempering~\cite{RoeverMeyerChristensen2007a} 
and yielded a sampler that reliably converged towards 
the true posterior distribution in the examples discussed below.
More details on the implementation of the evolutionary MCMC
algorithm can be found in section~\ref{sec:MCMCimpl}.

\subsection{Data and signal waveform}
Our simulated data consist of simultaneous measurements 
from several interferometric detectors, 
superimposed with interferometer-specific Gaussian noise.
The signal waveform that was injected into 
and recovered from the data
was implemented using a 3.5~post-Newtonian (PN)
approximation for the 
phase evolution~\cite{BlanchetEtAl2002,BlanchetEtAl2004},
and a 2.5~PN model for the amplitude~\cite{ArunEtAl2004}. 
The 9~parameters determining the responses at different interferometers are:
individual masses ($m_1, m_2 \in \realline^+;\; m_1\leq m_2$),
luminosity distance ($d_L \in \realline^+$),
inclination angle ($\iota \in [0,\pi]$),
coalescence phase ($\phi_0 \in [0,2\pi]$),
coalescence time at geocenter ($t_c \in \realline$),
declination ($\delta \in [-\frac{\pi}{2},\frac{\pi}{2}]$),
right ascension ($\alpha \in [0,2\pi]$) and
polarization angle ($\psi \in [0,\pi]$).
In order to derive the waveform at an individual interferometer~$I$,
the `local parameters'
altitude ($\vartheta^{(I)}$), azimuth ($\varphi^{(I)}$) 
local coalescence time ($t_c^{(I)}$) and
local polarisation ($\psi^{(I)}$) need to be determined 
for each interferometer.
More specific definitions
are given in~\cite{RoeverMeyerChristensen2007a,Blanchet2001}.
An appropriate way of dealing with the failure of the waveform
approximation shortly before coalescence still needs to be found.
For now we simply terminate the waveform as soon as
the innermost stable orbit \cite{BlanchetEtAl2002} is reached, 
or as the (3.5~PN) approximated orbital frequency starts decreasing.
The latter usually happens first; it is a non-physical effect 
that has also been noticed in other contexts
and gives an indication of the obvious failure 
of the waveform approximation
\cite{BuonannoEtAl2006,BakerEtAl2006}.
In the future numerical integration might be used
to extrapolate still further.
The same approach could be used with a wide range of waveforms;
in previous studies we have
used implementations of the 2.0 PN stationary-phase approximation
\cite{RoeverMeyerChristensen2006a},
and a 2.5~PN phase and 2.0~PN amplitude approximation
\cite{RoeverMeyerChristensen2007a}.
We are currently also working on an extension to the case of
spinning binaries, which entails the consideration of several
additional parameters.

\subsection{Priors}
We applied non-informative priors on the `geometrical' parameters 
that describe the inspiral event's location and orientation. 
Assuming that any direction and orientation is equally likely
(or none of these is a~priori `preferred'),
this leads to uniform priors for right ascension~$\alpha$, 
polarization angle~$\psi$ and coalescence phase~$\phi_0$, 
and to prior densities
\begin{equation} \textstyle
  f(\delta) = \frac{1}{2}\cos(\delta) 
  \qquad \mbox{and} \qquad
  f(\iota) = \frac{1}{2}\sin(\iota) 
\end{equation}
for declination~$\delta$ and inclination angle~$\iota$.
These also define the \textsl{Maximum Entropy} settings
for these parameters~\cite{Jaynes}.
The coalescence time~$t_c$ is assumed to be known in advance 
up to a certain accuracy from the detection pipeline 
that would in reality precede such an 
analysis~\cite{AmicoEtAl2003}.
For our demonstration purposes we set its prior to be uniform across 
$\pm 5$~ms around the true value (which of course is
known for simulated data);
using wider ranges only makes the search phase longer, 
due to the larger parameter space \cite{RoeverMeyerChristensen2007a}.
The prior for the masses ($m_1$, $m_2$) reflects the
distribution of the masses among 
binary inspirals, which could be based on 
observational evidence~\cite{Finn1994,KerkwijkEtAl1995} 
as well as theoretical considerations~\cite{BelczynskiEtAl2002,Janka2004}.
For now, we simply defined it as 
uniform across a range of 1--10~$\sunmass$. 
In principle, this type of search will be applicable for
component masses up to 20~$\sunmass$
(which corresponds to the low frequency sensitivity limit
for LIGO and Virgo).

\begin{figure}[ht]
  \begin{center}
    \includegraphics[width=15cm]{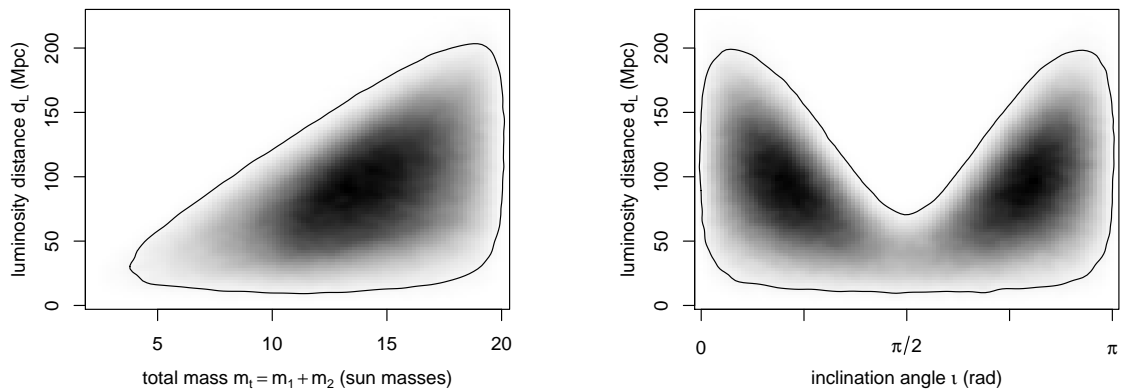}
    \caption{Marginal joint prior densities for total mass~$m_t$ and distance~$d_L$,
             and inclination~$\iota$ and distance~$d_L$. 
             The contour line encloses a~99\% credibility region.}
    \label{fig:priors}
  \end{center}
\end{figure}

Assuming that inspirals happen uniformly across space leads to 
a prior $\prob(d_L\leq x) \propto x^3$ for the luminosity
distance~$d_L$. This is an improper prior, seemingly implying 
there was an `infinite' probability for `infinitely remote' 
inspiral events. 
It is also unrealistic, since an inspiral event needs to happen
within a certain range in order to be detectable, 
otherwise its signal would be too faint to be noticed at all.
We incorporated this restriction into the prior specification
by considering the \textsl{detection probability} of an inspiral event, 
depending on the signal-to-noise ratio (SNR).
A signal's SNR increases linearly with its amplitude,
so for the prior definition we use the 
amplitude as an approximation to the SNR.
Further simplifying its expression (and considering only the 
intrinsic parameters' effects on the amplitude) we define
\begin{eqnarray} \label{eqn:ampli}
  \mathcal{A}(m_1, m_2, d_L, \iota) & := &
  \ln\Biggl(\frac{\sqrt{\eta}\;m_t^{\frac{5}{6}}}{d_L}\Biggr) 
  + \ln\biggl(\underbrace{\sqrt{(1+\cos(\iota)^2)^2 + (2\cos(\iota))^2}}_{\geq1\;\;\;\mathrm{and}\;\;\;\leq\sqrt{8}\,\approx\,2.8}\biggr) \nonumber \\
    & = & \ln\Biggl(\frac{\sqrt{\eta}\;m_t^{\frac{5}{6}}}{d_L}\Biggr) 
  + \frac{1}{2}\ln\Bigl(1+6\cos^2\iota+\cos^4\iota\Bigr),
\end{eqnarray}
where $m_t=m_1+m_2$ is the \textsl{total mass},
and $\eta=\frac{m_1m_2}{m_t^2}$ is the \textsl{(symmetric) mass ratio}
of the inspiralling system
($\mathcal{A}$ is actually proportional to 
the \textsl{logarithmic} amplitude)~\cite{RoeverMeyerChristensen2006a}.
Now one could set a threshold amplitude below which the corresponding
event would be considered undetectable, but we preferred a smoother transition
that doesn't strictly rule out parts of the parameter space. 
We do so by modeling the \textsl{detection probability}
\begin{equation}  \label{eqn:DetectProb}
  D_{a,b}(x) = \frac{1}{1+\exp(\frac{x-a}{b})}
\end{equation}
as a (sigmoidal) function of the amplitude value~$x$.
The values of~$a$ and~$b$ are set by defining at which amplitudes $x_L$ and $x_U$ 
the detection probability reaches some value~$p$ and exceeds $1-p$ 
(where $0<p<0.5$, e.g.\ $p:=0.1$). Given $x_L$ and $x_U$, 
these are set to
\begin{equation}  \label{eqn:SigmoidParam}
  a:= \frac{x_L+x_U}{2} \qquad \mbox{and} 
  \qquad 
  b:= \frac{x_U-x_L}{2\,\log(\frac{p}{1-p})}.
\end{equation}
In the following, we defined $p:=0.1$,
$x_U:=\mathcal{A}(2\sunmass,2\sunmass,50\mathrm{Mpc},0)$ and
$x_L:=\mathcal{A}(2\sunmass,2\sunmass,60\mathrm{Mpc},0)$,
assuming that a 2-2~$\sunmass$ inspiral with zero inclination 
is detectable out to distances of 50 and 60~Mpc with
90\% and 10\% probability, respectively,
and providing a reasonable coverage of the parameter space
for the example below.
More realistic bounds may be specified with respect to a certain
detection pipeline that is supposed to be installed upstream.
Considerations within a similar context 
(long-term observations of pulsars' gravitational wave signals)
indeed show that while detection of signals 
is certain for high amplitudes and impossible for low amplitudes, 
there also is a transition region in between 
where detectability is a matter of chance~\cite{Umstaetter2006,UmstaetterEtAl2007}.
The detection probability then enters the prior definition as an additional factor.
Considering `occurrence' and `detection' probabilities this way 
then leads to a proper prior distribution for all parameters,
reflecting the knowledge about the inspiral signal \textsl{given} that it
released a trigger in the pipeline.
Figure~\ref{fig:priors} illustrates some marginal prior densities 
resulting from the above settings.

One could actually explicitly use the SNR for the prior 
instead of approximating it by~$\mathcal{A}$. 
An SNR computation in general is computationally about as expensive
as a likelihood evaluation; but once one already did the likelihood 
evaluation for given parameter values, the SNR computation would simplify, 
since it could partly build on computations done in the first step.
When approximating the SNR by $\mathcal{A}$, any effects of the mass parameters
besides their effect on the overall amplitude are neglected,
as well as the impact of the antennae patterns, 
i.e.\ the sensitivities of the individual interferometers 
with respect
to the signal's sky location.

The above definitions imply that e.g.\ greater masses have a greater 
prior probability 
(since, resulting in greater amplitudes, 
they are still detectable at farther distances, see figure~\ref{fig:priors}),
although \textsl{initially} any masses were assumed to occur equally likely.
An analogous effect is known in astronomy as the 
\textsl{Malmquist effect};
incorporating it into the prior definition will compensate for
selection bias that would otherwise affect the 
parameter estimates~\cite{Teerikorpi1997,HendrySimmons1995}.

\subsection{Likelihood}
We assume that the noise is independent between different interferometers.
Con\-se\-quent\-ly, the likelihood for each site can be computed individually, 
and the network likelihood then arises as the product of the individual likelihoods.
Individual likelihoods are computed based on Fourier transforms 
of data and signal, and the noise spectrum, that is specific for 
each interferometer~\cite{FinnChernoff1993}. 
More details about the likelihood computation are given 
in~\cite{RoeverMeyerChristensen2007a}.

\subsection{MCMC implementation}\label{sec:MCMCimpl}
We implemented the MCMC sampler as a basic Metropolis 
algorithm~\cite{BDA,MCMCinPractice} that first was extended to a
parallel tempering algorithm.
The `tempering' here works as in a 
simulated annealing algorithm~\cite{NumRecipes}, and prevents 
MCMC chains from getting stuck in local modes of the posterior
distribution.
Parallel tempering then is the special case of a 
Metropolis-coupled MCMC (MCMCMC) algorithm~\cite{MCMCinPractice},
where several tempered MCMC chains, each at different temperatures, 
are run in parallel, and additional proposals are introduced 
to `swap' parameter sets between 
chains~\cite{HukushimaNemoto1996,RoeverMeyerChristensen2007a}.
This algorithm can be further refined by implementing elements
of genetic algorithms~\cite{Goldberg}. 
The set of parallel chains may be thought of as constituting a
`population' whose individuals may be crossed to form `hybrids'
that inherit properties from both `parental' chains,
the result being an evolutionary MCMC algorithm~\cite{LiangWong2001a}.
The `crossovers' between sets of parameters were implemented
as \textsl{real crossovers}, in which offsprings are formed 
by randomly reassembling the parental parameter sets, 
as well as \textsl{snooker crossovers},
in which a new offspring is proposed somewhere on the straight line 
connecting the two parental points in 
parameter space~\cite{GilksRobertsGeorge1994}.
Internally, instead of the original mass parameters, the
chirp mass~$m_c=\frac{(m_1 m_2)^{3/5}}{(m_1+m_2)^{1/5}}$
and the (symmetric) mass ratio~$\eta=\frac{m_1 m_2}{(m_1+m_2)^2}$
were used,
since these are easier to sample from.

\section{Example application}
We applied our MCMC routine to a simulated data set, 
corresponding to an inspiral signal that is received at three interferometers,
specifically the two LIGO sites Hanford (LHO) and Livingston (LLO), and the Virgo
interferometer near Pisa (V).
The simulated inspiral involved masses 
of $m_1=2\,\sunmass$ and $m_2=5\,\sunmass$
(chirp mass~$m_c=2.70\,\sunmass$, mass ratio~$\eta=0.204$),
observed from a distance of $d_L=30\,\mbox{Mpc}$
at $t_c=700\,009\,012.345$ GPS seconds. 
For the synthesized data that we use the noise characteristics were assumed to
match the target sensitivities for LIGO and Virgo~\cite{BeauvilleEtAl2005}.
The resulting SNRs~\cite{RoeverMeyerChristensen2007a}
at the three sites were 8.4~(LHO), 10.9~(LLO), 6.4~(V), 
and the network SNR was~15.2.

\begin{figure}[ht]
  \begin{center}
    \includegraphics[width=15cm]{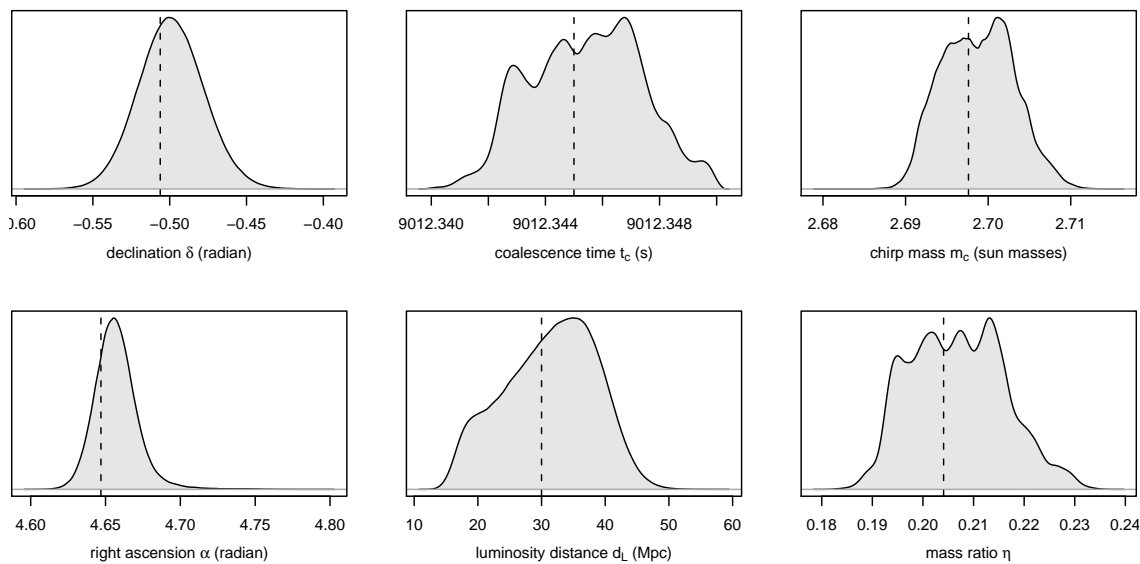}
    \caption{\label{fig:posterior1}Marginal joint posterior densities 
             for some of the parameters.
             Dashed lines indicate the true parameter values.}
  \end{center}
\end{figure}
\begin{figure}[b]
  \begin{center}
    \includegraphics[width=15cm]{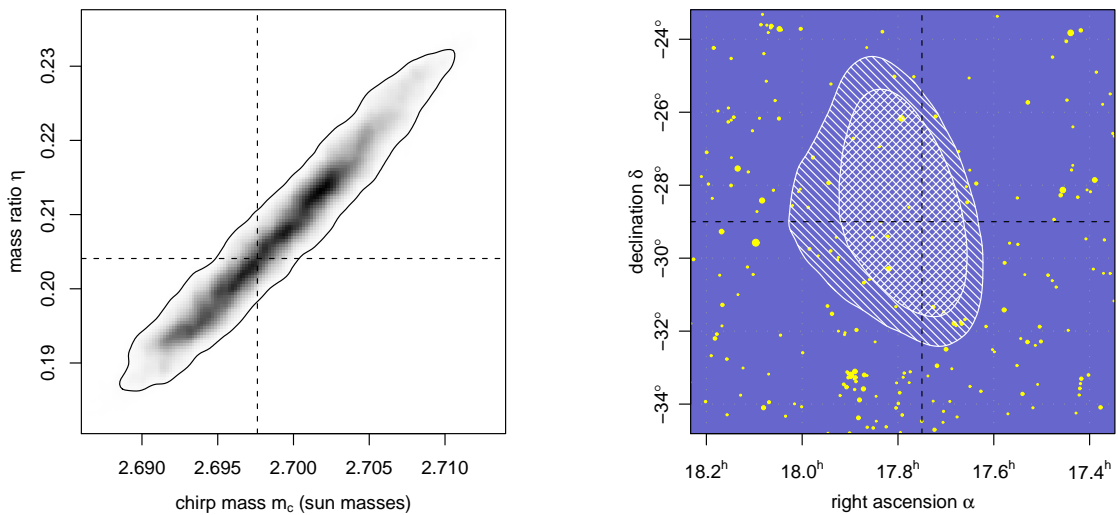}
    \caption{\label{fig:posterior2}Left: Marginal joint posterior density 
             for the two mass parameters, and a 99\% credibility region.
             Right: 95\% and 99\% credibility regions for the sky location
             against the backdrop of the night sky.
             Dashed lines indicate the true values.}
  \end{center}
\end{figure}

Figure~\ref{fig:posterior1} 
shows the marginal posterior distributions for several individual parameters
in comparison to the true values for the injected signal.
While some of the (marginal) distributions appear roughly Gaussian,
others are clearly not, and some even possess multiple modes. 
This illustrates some of the strengths of a fully Bayesian approach:
no approximations to the posterior's (or likelihood's) shape are made, 
an irregular posterior surface does not pose a problem, 
and the assessment of relative importance of multiple modes 
arises as a matter of course \cite{RoeverMeyerChristensen2007a}.
Figure~\ref{fig:posterior2} illustrates the joint distributions 
of two pairs of parameters.
The high correlation between chirp mass and mass ratio
indicates some degeneracy between these two parameters.
Table~\ref{tab:estimates} 
lists some numerical estimates for individual parameters for our
specific example.

\begin{table}
\caption{\label{tab:estimates}Some key figures of the individual parameters' 
         marginal posterior distributions, where meaningful. 
         Mean and standard deviation illustrate location and spread,
         and the 95\%~central credible interval gives a range that contains
         the true parameter with 95\%~probability, given the data at hand.}
\begin{center}
\small
\begin{tabular}{rcccccl} \br
       & mean &  st.dev. & 95\% c.c.i. & true & unit \\ \mr
 chirp mass ($m_c$)           & $2.6988$  & $0.0043$ &  ($2.6913$,  $ 2.7071$)   & $ 2.6976$ & $\sunmass$\\ 
 mass ratio ($\eta$)          & $0.2069$  & $0.0092$ &  ($0.1917$,  $ 0.2258$)   & $ 0.2041$ & \\
 coalescence time ($t_c$)     & $12.3454$ & $0.0019$ & ($12.3420$,  $12.3491$)   & $12.3450$ & s\\
 luminosity distance ($d_L$)  & $31.3$    & $ 7.2  $ & ($17.4$,     $43.6$)      & $30.0   $ & Mpc\\
 inclination angle ($\iota$)  & $0.737$   & $0.343 $ &  ($0.160$,   $ 1.462$)    & $ 0.700 $ & rad\\
 declination ($\delta$)       & $-0.499^\mathrm{a}$ & & ($-0.540$, $-0.457$) & $-0.506$  & rad\\
 right ascension ($\alpha$)   & $4.657^\mathrm{a}$ 
                              & \raisebox{1.5ex}[-1.5ex]{$\biggr\}\mbox{0.025}^\mathrm{a}\phantom{\biggr\}}$} 
                              & ($4.632$, $4.689$) & $4.647$ & rad\\
 coalescence phase ($\phi_0$) & $2.84^\mathrm{a}$ & $1.38^\mathrm{a}$ & & $2.0$    & rad\\
 \br
 \multicolumn{6}{l}{\footnotesize $^\mathrm{a}$ \textsl{mean direction} 
                    and \textsl{spherical st.dev}.\
                    (suitable for angular variables)
                    \cite{MardiaJupp}}
\end{tabular}
\end{center}
\end{table}

\section{Discussion}
We have presented a description of our coherent MCMC code for estimating nine
parameters associated with a binary inspiral signal detected by a network of
interferometric detectors. 
This program uses time-domain inspiral templates that
are 3.5~PN in phase and 2.5~PN in amplitude. 
New MCMC techniques, such as evolutionary MCMC and genetic algorithms, 
have been implemented in our code. 
The code can be applied to inspiral signals where the masses 
of the components can be as large as $20\;\sunmass$;
inspirals with large mass ratios can also be successfully analyzed. 
This code is part of a large mass ratio inspiral detection pipeline 
that we are currently developing; 
a \textsl{loose-net} inspiral detection pipeline (using, for example, lower order
PN templates) will generate a reasonable number of triggers, and this MCMC will
then be applied to those times where triggers were recorded. 
The next logical extension of our binary inspiral MCMC work will be to 
systems with spin. The addition of spin will increase the number of 
parameters needed for the model, 
and consequently will increase the complexity and the time 
required to run the MCMC. This is currently an area of active research 
for us.

\ack
This work was supported by the Marsden Fund Council from Government funding,
administered by the Royal Society of New Zealand (grant \mbox{UOA-204}), 
National Science Foundation grant \mbox{PHY-0553422}, and the Fulbright Scholar
Program.

\section*{References}
  \bibliographystyle{unsrt}
  \bibliography{/home/phd/christian/literature/literature}

\begin{thebibliography}{10}

\bibitem{AbbottEtAl2004a}
B.~Abbott et~al.
\newblock Detector description and performance for the first coincidence
  observations between {LIGO} and {GEO}.
\newblock {\em Nuclear Instruments and Methods in Physics Research~A},
  517:154--179, January 2004.

\bibitem{Sigg2004}
D.~Sigg.
\newblock Commissioning of {LIGO} detectors.
\newblock {\em Classical and Quantum Gravity}, 21(5):S409--S415, March 2004.

\bibitem{AcerneseEtAl2006}
F.~Acernese et~al.
\newblock The status of {VIRGO}.
\newblock {\em Classical and Quantum Gravity}, 23(8):S63--S70, April 2006.

\bibitem{AcerneseEtAl2005}
F.~Acernese et~al.
\newblock Status of {VIRGO}.
\newblock {\em Classical and Quantum Gravity}, 22(18):S869--S880, September
  2005.

\bibitem{LueckEtAl2006}
H.~L\"{u}ck et~al.
\newblock Status of the {GEO600} detector.
\newblock {\em Classical and Quantum Gravity}, 23(8):S71--S78, April 2006.

\bibitem{TakahashiEtAl2003}
R.~Takahashi et~al.
\newblock Operational status of {TAMA300}.
\newblock {\em Classical and Quantum Gravity}, 20(7):S593--S598, September
  2003.

\bibitem{TaylorWeisberg1989}
J.~H. Taylor and J.~M. Weisberg.
\newblock Further experimental tests of relativistic gravity using the binary
  pulsar {PSR1913+16}.
\newblock {\em The Astrophysical Journal}, 345:434--450, October 1989.

\bibitem{AbbottEtAl2004b}
B.~Abbott et~al.
\newblock Analysis of {LIGO} data for gravitational waves from binary neutron
  stars.
\newblock {\em Physical Review~{D}}, 69(12):122001, June 2004.

\bibitem{MarionEtAl2003}
F.~Marion et~al.
\newblock Gravitational waves and experimental gravity.
\newblock In {\em Proceedings of the Rencontres de Moriond~2003}. Editions
  Fronti\'{a}eres, Gif-sur-Yvette, France, 2004.

\bibitem{AmicoEtAl2003}
P.~Amico et~al.
\newblock A parallel {B}eowulf-based system for the detection of gravitational
  waves in interferometric detectors.
\newblock {\em Computer Physics Communications}, 153(2):179--189, June 2003.

\bibitem{BeauvilleEtAl2007}
F.~Beauville et~al.
\newblock Detailed comparison of {LIGO} and {V}irgo inspiral pipelines in
  preparation for a joint search.
\newblock {\em Arxiv preprint gr-qc/0701027}, January 2007.

\bibitem{AbbottEtAl2005a}
B.~Abbott et~al.
\newblock Search for gravitational waves from galactic and extra-galactic
  binary neutron stars.
\newblock {\em Physical Review~{D}}, 72(8):082001, October 2005.

\bibitem{AbbottEtAl2005b}
B.~Abbott et~al.
\newblock Search for gravitational waves from primordial black hole binary
  coalescences in the galactic halo.
\newblock {\em Physical Review~{D}}, 72(8):082002, October 2005.

\bibitem{AbbottEtAl2006a}
B.~Abbott et~al.
\newblock Search for gravitational waves from binary black hole inspirals in
  {LIGO} data.
\newblock {\em Physical Review~{D}}, 73(6):062001, March 2006.

\bibitem{AbbottEtAl2006b}
B.~Abbott et~al.
\newblock Joint {LIGO} and {TAMA300} search for gravitational waves from
  inspiralling neutron star binaries.
\newblock {\em Physical Review~{D}}, 73(10):102002, May 2006.

\bibitem{MCMCinPractice}
W.~R. Gilks, S.~Richardson, and D.~J. Spiegelhalter.
\newblock {\em Markov chain Monte Carlo in practice}.
\newblock Chapman \& Hall / CRC, Boca Raton, 1996.

\bibitem{RoeverMeyerChristensen2006a}
C.~R\"{o}ver, R.~Meyer, and N.~Christensen.
\newblock Bayesian inference on compact binary inspiral gravitational radiation
  signals in interferometric data.
\newblock {\em Classical and Quantum Gravity}, 23(15):4895--4906, August 2006.

\bibitem{RoeverMeyerChristensen2007a}
C.~R\"{o}ver, R.~Meyer, and N.~Christensen.
\newblock Coherent {B}ayesian inference on compact binary inspirals using a
  network of interferometric gravitational wave detectors.
\newblock {\em Physical Review~{D}}, 75(6):062004, March 2007.

\bibitem{BlanchetEtAl2002}
L.~Blanchet, G.~Faye, B.~R. Iyer, and B.~Joguet.
\newblock Gravitational-wave inspiral of compact binary systems to
  7/2~post-{N}ewtonian order.
\newblock {\em Physical Review~{D}}, 65(6):061501, March 2002.
\newblock Note the erratum \cite{BlanchetEtAl2005a}.

\bibitem{BlanchetEtAl2005a}
L.~Blanchet, G.~Faye, B.~R. Iyer, and B.~Joguet.
\newblock Erratum: Gravitational-wave inspiral of compact binary systems to
  7/2~post-{N}ewtonian order.
\newblock {\em Physical Review~{D}}, 71(12):129902, June 2005.
\newblock (See also \cite{BlanchetEtAl2002}).

\bibitem{ArunEtAl2004}
K.~G. Arun, L.~Blanchet, B.~R. Iyer, and M.~S.~S. Qusailah.
\newblock The 2.5~{PN} gravitational wave polarizations from inspiralling
  compact binaries in circular orbits.
\newblock {\em Classical and Quantum Gravity}, 21(15):3771--3801, August 2004.
\newblock Note the erratum \cite{ArunEtAl2005}.

\bibitem{ArunEtAl2005}
K.~G. Arun, L.~Blanchet, B.~R. Iyer, and M.~S.~S. Qusailah.
\newblock Corrigendum: The 2.5{PN} gravitational wave polarizations from
  inspiralling compact binaries in circular orbits.
\newblock {\em Classical and Quantum Gravity}, 22(14):3115--3117, July 2005.
\newblock (See also \cite{ArunEtAl2004}).

\bibitem{LiangWong2001a}
F.~Liang and H.~W. Wong.
\newblock Real-parameter {E}volutionary {M}onte {C}arlo with applications to
  {B}ayesian mixture models.
\newblock {\em Journal of the American Statistical Association},
  96(454):653--666, June 2001.

\bibitem{Finn1997}
L.~S. Finn.
\newblock Issues in gravitational wave data analysis.
\newblock {\em Arxiv preprint gr-qc/9709077}, September 1997.

\bibitem{Jaynes}
E.~T. Jaynes.
\newblock {\em Probability theory: The logic of science}.
\newblock Cambridge University Press, Cambridge, 2003.

\bibitem{Gregory}
P.~C. Gregory.
\newblock {\em Bayesian logical data analysis for the physical sciences}.
\newblock Cambridge University Press, Cambridge, 2005.

\bibitem{BDA}
A.~Gelman, J.~B. Carlin, H.~Stern, and D.~B. Rubin.
\newblock {\em Bayesian data analysis}.
\newblock Chapman \& Hall / CRC, Boca Raton, 1997.

\bibitem{NumRecipes}
W.~H. Press, S.~A. Teukolsky, W.~T. Vetterling, and B.~P. Flannery.
\newblock {\em Numerical recipes in~{C}: The art of scientific computing}.
\newblock Cambridge University Press, Cambridge, 2nd edition, 1992.

\bibitem{Goldberg}
D.~E. Goldberg.
\newblock {\em Genetic algorithms in search, optimization, and machine
  learning}.
\newblock Addison-Wesley, Reading, Mass., 1989.

\bibitem{BlanchetEtAl2004}
L.~Blanchet, T.~Damour, G.~Esposito-Far\`{e}se, and B.~R. Iyer.
\newblock Gravitational radiation from inspiralling compact binaries completed
  at the third post-{N}ewtonian order.
\newblock {\em Physical Review Letters}, 93(9):091101, August 2004.

\bibitem{Blanchet2001}
L.~Blanchet.
\newblock Post-{N}ewtonian computation of binary inspiral waveforms.
\newblock In I.~Ciufolini, V.~Gorini, U.~Moschella, and P.~Fr\'{e}, editors,
  {\em Gravitational waves: Proceedings of the {C}omo school on gravitational
  waves in astrophysics}. Institute of Physics Publishing, Bristol, 2001.
\newblock See also Arxiv preprint gr-qc/0104084.

\bibitem{BuonannoEtAl2006}
A.~Buonanno, G.~B. Cook, and F.~Pretorius.
\newblock Inspiral, merger and ring-down of equal-mass black-hole binaries.
\newblock {\em Arxiv preprint gr-qc/0610122}, October 2006.

\bibitem{BakerEtAl2006}
J.~G. Baker, J.~R. van Meter, S.~T. McWilliams, J.~Centrella, and Kelly~B. J.
\newblock Consistency of post-{N}ewtonian waveforms with numerical relativity.
\newblock {\em Arxiv preprint gr-qc/0612024}, December 2006.

\bibitem{Finn1994}
L.~S. Finn.
\newblock Observational constraints on the neutron star mass distribution.
\newblock {\em Physical Review Letters}, 73(14):1878--1881, October 1994.

\bibitem{KerkwijkEtAl1995}
M.~H. van Kerkwijk, J.~van Paradijs, and E.~J. Zuiderwijk.
\newblock On the masses of neutron stars.
\newblock {\em Astronomy and Astrophysics}, 303:497--501, 1995.

\bibitem{BelczynskiEtAl2002}
K.~Belczynski, V.~Kalogera, and T.~Bulik.
\newblock A comprehensive study of binary compact objects as gravitational wave
  sources: evolutionary channels, rates, and physical properties.
\newblock {\em The Astrophysical Journal}, 572(1):407--431, June 2002.

\bibitem{Janka2004}
H.-T. Janka.
\newblock Neutron star formation and birth properties.
\newblock {\em Arxiv preprint astro-ph/0402200}, February 2004.

\bibitem{Umstaetter2006}
R.~Umst\"{a}tter.
\newblock {\em Bayesian strategies for gravitational radiation data analysis}.
\newblock PhD thesis, The University of Auckland, 2006.
\newblock URL \url{http://hdl.handle.net/2292/377}.

\bibitem{UmstaetterEtAl2007}
R.~Umst\"{a}tter et~al.
\newblock Setting upper limits from {LIGO} on gravitational waves from
  {SN1987a}.
\newblock {\em Preprint in preparation}, 2007.

\bibitem{Teerikorpi1997}
P.~Teerikorpi.
\newblock Observational selection bias affecting the determination of the
  extragalactic distance scale.
\newblock {\em Annual Review of Astronomy and Astrophysics}, 35:101--136,
  September 1997.

\bibitem{HendrySimmons1995}
M.~A. Hendry and J.~F.~L. Simmons.
\newblock Distance estimation in cosmology.
\newblock {\em Vistas in Astronomy}, 39(3):297--314, 1995.

\bibitem{FinnChernoff1993}
L.~S. Finn and D.~F. Chernoff.
\newblock Observing binary inspiral in gravitational radiation: One
  interferometer.
\newblock {\em Physical Review~{D}}, 47(6):2198--2219, March 1993.

\bibitem{HukushimaNemoto1996}
K.~Hukushima and K.~Nemoto.
\newblock Exchange {M}onte {C}arlo method and application to {S}pin {G}lass
  simulations.
\newblock {\em Journal of the Physical Society of Japan}, 65(6):1604--1608,
  June 1996.

\bibitem{GilksRobertsGeorge1994}
W.~R. Gilks, G.~O. Roberts, and E.~I. George.
\newblock Adaptive direction sampling.
\newblock {\em The Statistician}, 43(1):179--189, 1994.

\bibitem{BeauvilleEtAl2005}
F.~Beauville et~al.
\newblock A first comparison between {LIGO} and {V}irgo inspiral search
  pipelines.
\newblock {\em Classical and Quantum Gravity}, 22(18):S1149--S1158, September
  2005.

\bibitem{MardiaJupp}
K.~V. Mardia and P.~E. Jupp.
\newblock {\em Directional statistics}.
\newblock Wiley \& Sons, Chichester, 2000.

\end{thebibliography}

\end{document}